\documentclass[twoside,12pt]{article}
\usepackage{CJKutf8}
\usepackage{indentfirst}
\usepackage{bm}
\usepackage{epsfig}
\usepackage{graphicx}
\usepackage{subfig}
\usepackage{booktabs}
\usepackage{caption}
\usepackage{multirow}
\usepackage{amssymb,amsmath}
\usepackage{cite}
\usepackage{color} 
\usepackage[section]{placeins}

\begin{document}
\begin{CJK}{UTF8}{gbsn}

\thispagestyle{empty} \vspace*{0.8cm}\hbox
to\textwidth{\vbox{\hfill\huge\sf Commun. Theor. Phys.\hfill}}
\par\noindent\rule[3mm]{\textwidth}{0.2pt}\hspace*{-\textwidth}\noindent
\rule[2.5mm]{\textwidth}{0.2pt}


\begin{center}
\LARGE\textbf{Dark energy effects on surface gravitational redshift and Keplerian frequency of neutron stars}
\end{center}

\footnotetext{\hspace*{-.45cm}\footnotesize $^*...........$.}
\footnotetext{\hspace*{-.45cm}\footnotesize $^\dag$ E-mail: {\it xuy@cho.ac.cn}  }

\begin{center}
\rm Jia-Jing He$^{\rm 1,2}$, \ \ Yan Xu$^{\rm 1,2,\dagger}$, \ \ Yi-Bo Wang$^{\rm 1}$, \ \ Xiu-Lin Huang$^{\rm 1}$, \ \ Xing-Xing Hu$^{\rm 1}$ and  \ Yu-Fu Shen $^{\rm 1}$
\end{center}

\begin{center}
\begin{footnotesize} \sl
${}^{\rm 1}$ Changchun Observatory, National Astronomical Observatories, Chinese Academy of Sciences, Changchun 130117, China  \\
${}^{\rm 2}$ School of Astronomy and Space Sciences, University of Chinese Academy of Sciences, Beijing 100049, China\\
\end{footnotesize}
\end{center}

\begin{center}
\footnotesize (Received XXXX; revised manuscript received XXXX)

\end{center}

\vspace*{2mm}

\begin{center}
\begin{minipage}{15.5cm}
\parindent 20pt\footnotesize
The research of the properties of neutron stars with dark energy is a particularly interesting yet unresolved problem in astrophysics. We analyze the influence of dark energy on the equation of state, the maximum mass, the surface gravitational redshift and the Keplerian frequency for the traditional neutron star and the hyperon star matter within the relativistic mean field theory, using the GM1 and TM1 parameter sets by considering the two ﬂavor symmetries of SU(6) and SU(3) combined with the observations of PSR J1614-2230, PSR J0348+0432, PSR J0030+0451, RX J0720.4-3125, and 1E 1207.4-5209. It is found that the existence of dark energy leads to the softened equations of the state of the traditional neutron star and the hyperon star. The radius of a fixed-mass traditional neutron star (or hyperon star) with dark energy becomes smaller, which leads to the increased compactness. The existence of dark energy can also enhance the surface gravitational redshift and the Keplerian frequency of the traditional neutron stars and the hyperon stars. The growth of the Keplerian frequency may cause speeding up to the spin rate which may provide a possible way for understanding and explaining the pulsar glitch phenomenon. Specifically, we infer that the mass and the surface gravitational redshift of PSR J1748-2446ad without dark energy for the GM1 (TM1) parameter set are 1.141 $M_\odot$ (1.309 $M_\odot$) and 0.095 (0.105), respectively. The corresponding values for the GM1 (TM1) parameter set are 0.901 $M_\odot$ (1.072$M_\odot$) and 0.079 (0.091) if PSR J1748-2446ad contains dark energy with $\alpha=0.05$. PSR J1748-2446ad  may be a low-mass pulsar with a lower surface gravitational redshift under our selected models. 

\end{minipage}
\end{center}

\begin{center}
\begin{minipage}{15.5cm}
\begin{minipage}[t]{2.3cm}{\bf Keywords:}\end{minipage}
\begin{minipage}[t]{13.1cm}
dark energy; neutron stars; gravitational redshift; Keplerian frequency 
\end{minipage}\par\vglue8pt

\end{minipage}
\end{center}

\section{Introduction}
Ordinary matter occupies only about 5\%, while dark matter and dark energy account for approximately 25\% and 70\% of the cosmic space, respectively\textsuperscript{\cite{hinshaw2013nine,aghanim2020planck,abdullah2023constraining,ostriker2003new}}. Dark energy is primarily proposed to interpret the accelerating expansion of the universe\textsuperscript{\cite{riess1998observational,perlmutter1999measurements}}, which is regarded as a form of energy that pervades the universe and is characterized by a negative pressure. How to understand the physical properties of dark energy has become a fundamental and urgent issue in the fields of astronomy and physics\textsuperscript{\cite{huterer2017dark,frusciante2020effective,li2024prospects,jin2023prospects}}. The exploration and research of dark energy are crucial not only to help us comprehend the laws of the cosmic evolution, but also serve as a significant impetus for promoting scientific progress. 

Generally, the theoretical models of dark energy are studied in the context of cosmology. It is very beneficial to take the potential impact of dark energy on the compact stars into consideration. And it is generally believed that there is a non-coupling or very small coupling interaction between dark energy and baryonic matter. Thus the theoretical numerical simulation based on the pulsar observations is a good way to gain further understanding and to explore the influence of dark energy on the properties of neutron stars. Many meaningful works for neutron stars mixing dark energy have been done based on the different models\textsuperscript{\cite{yazadjiev2012possible,hess2015neutron,banerjee2020analytical,smerechynskyi2021impact,dayanandan2021modelling,sagar2022gravitationally,astashenok2023compact,tudeshki2024effect,pretel2024normal}}. For example, the oscillation spectra of neutron stars containing the different proportions of ordinary matter against dark energy were studied by Yazadjiev and Doneva in 2012\textsuperscript{\cite{yazadjiev2012possible}}. They pointed out that the fraction of dark energy had a significant impact on the frequencies of the fundamental mode and the higher overtones. How dark matter affects the macroscopic characteristics and the maximum mass of neutron stars by calculating the density distribution of dark energy was studied by Smerechynskyi et al. in 2021\textsuperscript{\cite{smerechynskyi2021impact}}. The structure of neutron stars with dark energy using the Tolman IV type gravitational potential was studied by Dayanandan and Smitha in 2022\textsuperscript{\cite{dayanandan2021modelling}}. The properties of hybrid stars including dark energy adopting the Tolman Buchdahl metric potential was researched by Sagar et al. in 2022\textsuperscript{\cite{sagar2022gravitationally}}. The mass-radius relationships between neutron stars and strange stars with dark energy based on General Relativity and Modified Gravity were researched by Astashenok et al. in 2023\textsuperscript{\cite{astashenok2023compact}}. 

It is well known that the measurement of pulsar mass is very difficult, and the situation of the measurement of pulsar radius is even worse. A total of 136 pulsars with the mass measurements was updated by Fan et al. in 2024\textsuperscript{\cite{fan2024maximum}}. Therefore, it is hoped that certain relationships between the mass and radius of pulsars. The gravitational redshift of the surface spectrum of the emission lines and the Keplerian frequency describing the maximum frequency of the stably rotating pulsars are exactly related to both the mass and radius. If the astronomical observations can simultaneously obtain the measurements of the mass and the gravitational redshift or the frequency of a pulsar, and then its radius can be estimated. Therefore, the surface gravitational redshift and the frequency of pulsars are often used to constrain the equations of state (EOSs) of neutron stars. However, there are very limited researches on the surface gravitational redshift and the Keplerian frequency of neutron star with dark energy. 

In the work, we studied the models of the traditional neutron stars and hyperon stars which are composed of dense matter and dark energy. We describe the dense matter under the framework of relativistic mean field theory and take the simple form from reference \cite{astashenok2023compact} as an example to describe dark energy, in which dark energy is considered as a simple fluid form. We will mainly explore the effects of dark energy on such as the EOS, the mass, the radius, the surface gravitational redshift and the Keplerian frequency of the traditional neutron star and the hyperon star. Natural units $c = G = 1$ are used throughout the work.


\section{Theory framework}
\subsection{The EOS and Tolman–Oppenheimer–Volkoff equations of neutron star mixing dark energy}


For the normal neutron star matter, the total Lagrangian density containing $\sigma$, $\omega$, $\rho$, $\sigma^{*}$ and $\phi$ mesons reads\textsuperscript{\cite{yaosong1999relativistic,miyatsu2013equation,yu2018effects}},
\begin{equation}
\begin{split}
\mathcal{L} & =\overline{\psi}_B[\mathrm{i}\gamma_\mu\partial^\mu-(m_B-g_{\sigma B}\sigma-g_{\sigma^*B}\sigma^*)-g_{\rho B}\gamma_{\mu}{\vec{\mathbf{\rho}}^\mu}\cdot\vec{I}_{B}-g_{\omega B}\gamma_\mu\omega^\mu-g_{\phi B}\gamma_\mu\phi^\mu]\psi_B\\
 &+\overline{\psi}_l[\mathrm{i}\gamma_\mu\partial^\mu-m_l]\psi_l
-\frac{1}{4}W^{\mu v}W_{\mu v}-\frac{1}{4}\vec{R}^{\mu v}\cdot\vec{R}_{\mu v}-\frac{1}{4}P^{\mu v}P_{\mu v} +\frac{1}{2}m_\omega^2\omega_\mu\omega^\mu\\
&+\frac{1}{2}m_\rho^2{\vec{\mathbf{\rho}}}_\mu\cdot{\vec{\mathbf{\rho}}}^\mu
+\frac{1}{2}m^2_{\phi}\phi_\mu\phi^\mu +\frac{1}{2}(\partial_\mu\sigma\partial^\mu\sigma-m_\sigma^2\sigma^2)
+\frac{1}{2}(\partial_v\sigma^*\partial^v\sigma^*-m^2_{\sigma^*}\sigma^{*2})\\
&-\frac{1}{3}g_2\sigma^{3}-\frac{1}{4}g_3\sigma^4+\frac{1}{4}c_3(\omega_\mu\omega^\mu)^2,      
\end{split}
\end{equation}

\noindent where $W_{\mu v}=\partial_\mu\omega_v-\partial_v\omega_\mu$, $R_{\mu v}=\partial_\mu{\mathbf{\rho}}_v-\partial_v{\mathbf{\rho}}_\mu$ and $P_{\mu v}=\partial_\mu\phi_v-\partial_v\phi_\mu$ are the field strength tensors of $\omega$, $\rho$ and $\phi$, respectively. The equations of motion of baryons and mesons could be derived by bringing Eq.(1) into the Euler-Lagrange equation. And then the energy density $\rho_m$ and the pressure $p_m$ of normal neutron star matter can be deduced as follows,
\begin{align}
\rho_{m} &=\sum_B\frac{1}{\pi^{2}}\int_0^{p_{FB}}\sqrt{p_{B}^{2}+m_{B}^{*2}}p_{B}^{2}dp_{B}+\frac{1}{3}g_2\sigma^{3}+\frac{1}{4}g_3\sigma^4 \nonumber\\
&+\frac{1}{2}m_{\sigma}^{2}\sigma^{2}+\frac{1}{2}m_{\sigma^{*}}^{2}\sigma^{*2}+\frac{1}{2}m_{\omega}^{2}\omega^{2}+\frac{3}{4}c_{3}\omega^{4}+\frac{1}{2}m_{\rho}^{2}\rho^{2} \nonumber\\
&+\frac{1}{2}m_{\phi}^{2}\phi^{2}+\sum_l\frac{1}{\pi^{2}}\int_0^{p_{Fl}}\sqrt{p_{l}^{2}+m_{l}^{*2}}p_{l}^{2}\mathrm{d}p_{l},\\
p_{m}&=\frac{1}{3}\sum_B\frac{1}{\pi^{2}}\int_0^{p_{FB}}\frac{p_{B}^{4}dp_{B}}{\sqrt{p_{B}^{2}+m_{B}^{*2}}}-\frac{1}{3}g_2\sigma^{3}-\frac{1}{4}g_3\sigma^4 \nonumber \\
&-\frac{1}{2}m_{\sigma}^{2}\sigma^{2}-\frac{1}{2}m_{\sigma^{*}}^{2}\sigma^{*2}+\frac{1}{2}m_{\omega}^{2}\omega^{2}+\frac{1}{4}c_{3}\omega^{4}+\frac{1}{2}m_{\rho}^{2}\rho^{2}\nonumber\\
&+\frac{1}{2}m_{\phi}^{2}\phi^{2}+\frac{1}{3}\sum_l\frac{1}{\pi^{2}}\int_0^{p_{Fl}}\frac{p_{l}^{4}dk}{\sqrt{p_{l}^{2}+m_{l}^{*2}}p_{l}^{2}}\mathrm{d}p_{l}.   
\end{align}

In the work, we choose TM1 and GM1 parameter sets and adpot two symmetries of SU(3) and SU(6) for the meson-hyperon couplings of relativistic mean field theory, which are listed in Table 1\textsuperscript{\cite{miyatsu2013equation}}.

If dark energy is assumed to be smooth in space, its contribution to the EOS based on the minimally-coupled scalar ﬁeld model of dark energy can be written as,
\begin{equation}
 p_{\mathrm{de}} = w\rho_{\mathrm{de}},    
\end{equation}
where $p_{de}$ and $\rho_{de}$ are the pressure and the energy density of dark energy, respectively. The values of the parameter $w$ are distinct in different models of dark energy. For example, for the cosmological constant model, $w=-1$. For the quintessence model, $-1<w< -\frac{1}{3}$\textsuperscript{\cite{wang2000cosmic}}. And for the phantom scalar field, $w<-1$\textsuperscript{\cite{caldwell2002phantom}}. In the work, we take $w \approx -1$ in that it fits well with the current observed data\textsuperscript{\cite{jassal2010understanding,mazumdar2021evidence}} and the Eq.(4) is given as,
\begin{equation}
    p_{\mathrm{de}}\approx -\rho_{\mathrm{de}},
\end{equation}
due to the unknown coupling between the neutron star matter and dark energy, we refer to the form of \cite{astashenok2023compact},
\begin{equation}
    \rho_{\mathrm{de}}=\alpha\rho_m {\mathrm{e}}^{-\frac{\rho_s}{\rho_m}},
\end{equation}
where the cutoff density $\rho_s$ is 300$ \rm MeV/{fm}^3$ and $\alpha$ is a positive constant. And we will gradually study more complex situations in the different dark energy models for neutron stars mixing dark energy in the future. 

The total energy density $\rho$ and the pressure $p$ are composed of two parts,
\begin{equation}
    p=p_m+p_{\mathrm{de}}, \hspace{2mm}     
    \rho=\rho_m+\rho_{\mathrm{de}}.
\end{equation}

The mass and radius of neutron star mixing dark energy can be obtained by solving the Tolman–Oppenheimer–Volkoff (TOV) equations for an anisotropic distribution,  
\begin{equation}
   \frac{\mathrm{d}p}{\mathrm{d}r}=-\frac{4\pi r^{3} p+m(r) }{r(r-2m(r))}(p+\rho),
\end{equation}
\begin{equation}
    \frac{\mathrm{d}{m(r)}}{\mathrm{d}r}= 4\pi r^{2}\rho.
\end{equation}
The Eqs.(8)-(9) satisfy the boundary conditions $m(r=0)=0$, $p(r=0)=p_c$ and $p(r=R)=0$ ($R$ is the star radius). 


\subsection{Gravitational redshift}
The surface spectrum of a neutron star shifts towards the red waveband due to the need to overcome the gravitational field when photons escape from the surface\textsuperscript{\cite{lackey2006observational}}. The phenomenon is known as the surface gravitational redshift which can be calculated by:
\begin{equation}
    Z=(1-\frac{2M}{R})^{-\frac{1}{2}}-1.
\end{equation}
It can be seen that the surface gravitational redshift can be obtained by solving the mentioned TOV equations.

\subsection{Keplerian frequency}

The Keplerian frequency is the maximum frequency at which a star can rotate stably. Once the rotational rate exceeds the Keplerian frequency, the material on the surface of the neutron star will be ejected due to the centrifugal force. In 2004, Lattimer and Prakash proposed a universal relation that does not rely on the EOS\textsuperscript{\cite{lattimer2004physics}}:
\begin{equation}
    f_K \approx C(\frac{M}{M_\odot})^{\frac{1}{2}}(\frac{R}{10km})^{-\frac{3}{2}}.
\end{equation}
When $C=1.08$ kHz, Eq.(11) can well describe neutron stars\textsuperscript{\cite{haensel2009keplerian}}. Similar to the surface gravitational redshift, the Keplerian frequency can be obtained by solving the TOV equations.

\section{The macroscopic properties and numerical analysis of neutron stars mixing dark energy}

In this work, we examine both the traditional neutron star and the hyperon star matter, where the former comprises neutrons, protons, electrons and muons, and the latter additionally includes hyperons. The baryonic matter and dark energy are characterized under the GM1 and TM1 parameter sets involving the two ﬂavor symmetries of SU(6) and SU(3). The numerical results are discussed and analyzed in the four situations with a positive constant $\alpha$ = 0, 0.025 and 0.05: (a) only contain nucleon (n, p) and lepton (e, $\mu$), namely, the traditional neutron star matter, (b) involve $\sigma$, $\omega$, $\rho$, $\sigma^{\ast}$ and $\phi$ mesons for describing the hyperon star matter under the SU(3) flavor symmetry, (c) involve $\sigma$, $\omega$, $\rho$, $\sigma^{\ast}$ and $\phi$ mesons for describing the hyperon star matter under the SU(6) spin-flavor symmetry, (d) only the nonstrange mesons ($\sigma$, $\omega$ and $\rho$) are included to describe the hyperon star matter under the SU(6) spin-flavor symmetry. The coupling constants of SU(3) and SU(6) flavor symmetries in GM1 and TM1 parameter sets are listed in Table 1. We will clarify the influence of dark energy on the EOS, the mass, the radius, the surface gravitational redshift and the Keplerian frequency of the traditional neutron star and the hyperon star, respectively.

\begin{table}[htb]
\tabcolsep=15pt\small
\centering
\captionsetup{labelfont=bf,labelsep=space}
\caption{The GM1 and TM1 parameter sets\textsuperscript{\cite{yaosong1999relativistic,miyatsu2013equation,yu2018effects}}.}
\begin{tabular}{ccccc}
\hline
Parameter Sets  & \multicolumn{2}{c}{GM1} & \multicolumn{2}{c}{TM1}  \\ \cmidrule(r){2-3} \cmidrule(r){4-5}
Symmetries & SU(3) & SU(6)& SU(3) & SU(6) \\
\hline
$g_{\sigma N}$& 9.57 & 9.57 & 10.029 & 10.029  \\
$g_2 ({fm}^{-1})$ &12.28 & 12.28 & 7.233 & 7.233 \\
$g_3$ & -8.98& -8.98 & 0.618 & 0.618  \\
$c_3$  & -& - & 81.601 & 71.308  \\
$g_{\omega N}$ &10.26 &  10.61 & 12.199& 12.614 \\
$g_{\rho N}$  &4.10 &  4.10 & 4.640 & 4.632  \\
$g_{\phi N}$ & -3.50 &- & -4.164& -  \\
$g_{\sigma\Lambda}$ & 7.25 & 5.84& 7.733&6.170  \\
$g_{\sigma\Sigma}$  & 5.28&  3.87& 6.035 & 4.472  \\
$g_{\sigma\Xi}$ & 5.87 &  3.06 & 6.328& 3.202  \\
$g_{\sigma^{\ast} N}$ & 0 &  0 & 0 & 0\\ 
$g_{\sigma^{\ast}\Lambda}$ & 2.60 &  3.73 & 3.691 & 5.015   \\
$g_{\sigma^{\ast}\Sigma}$ & 2.60 &  3.73 & 3.691 & 5.015 \\
$g_{\sigma^{\ast}\Xi}$  & 6.82 & 9.67 & 8.100 & 11.516\\
\hline \end{tabular} \small
\end{table}
\vspace*{2mm}

\begin{figure}[htb]%
    \centering
    \captionsetup{labelfont=bf,labelsep=period}
    \subfloat{
        \includegraphics[width=3.05 in,height=2.65in]{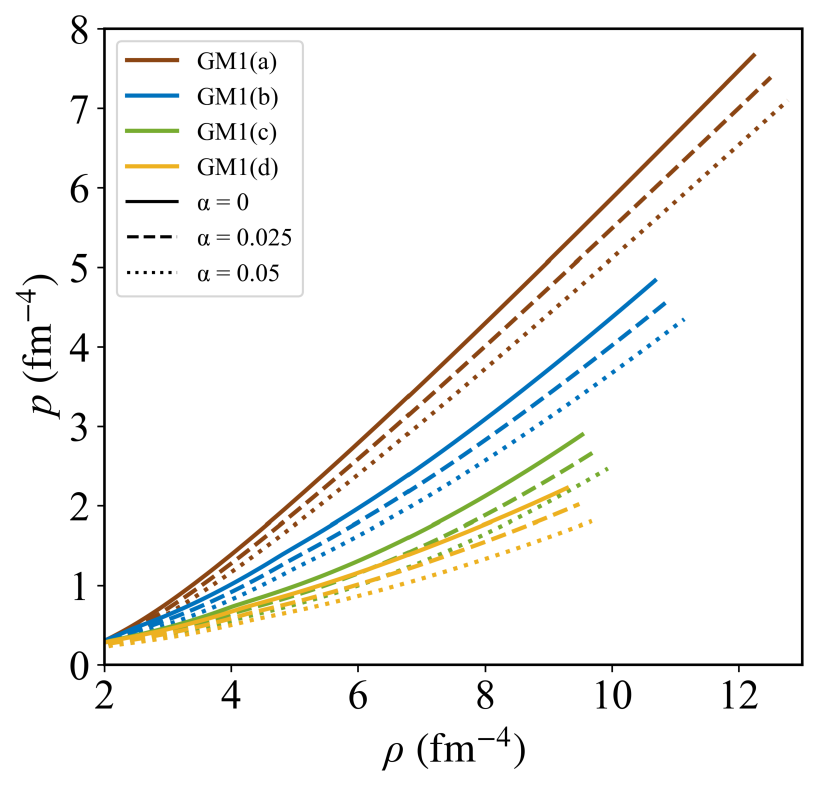}
        }\hspace{-1mm}
    \subfloat{
        \includegraphics[width=3.2 in,height=2.6in]{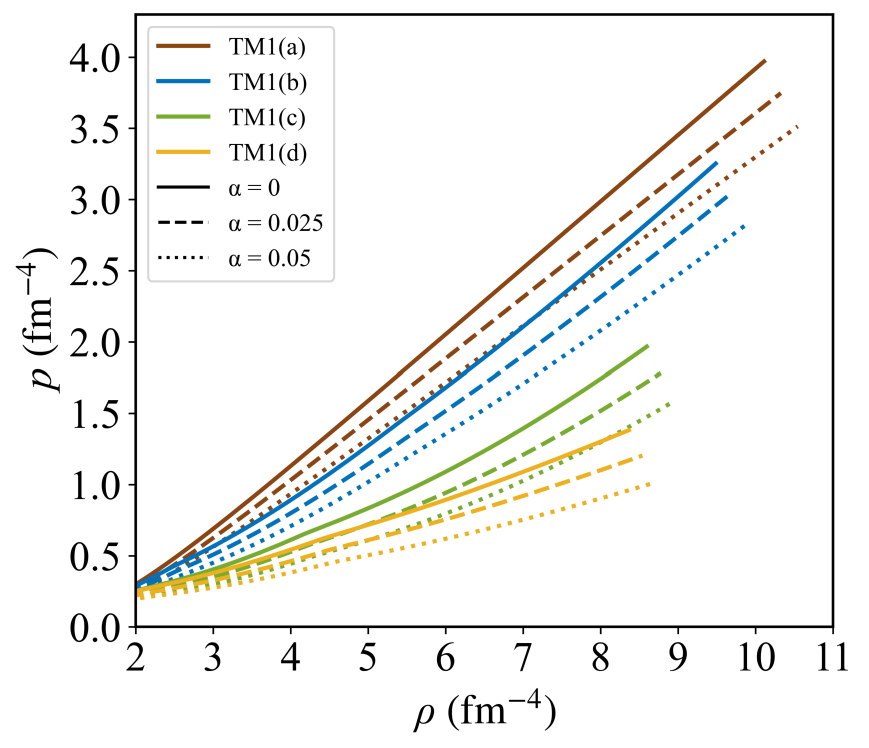}
        }
    \caption{EOSs of the traditional neutron stars and the hyperon stars under the GM1 and TM1 parameter sets. Solid, dashed and dotted lines correspond to the different fraction of dark energy with $\alpha$ = 0, 0.025 and 0.05, respectively.}
\end{figure}

\begin{figure}[htp]%
    \centering
    \captionsetup{labelfont=bf,labelsep=period}
    \subfloat{
        \includegraphics[width=3.2 in,height=2.8in]{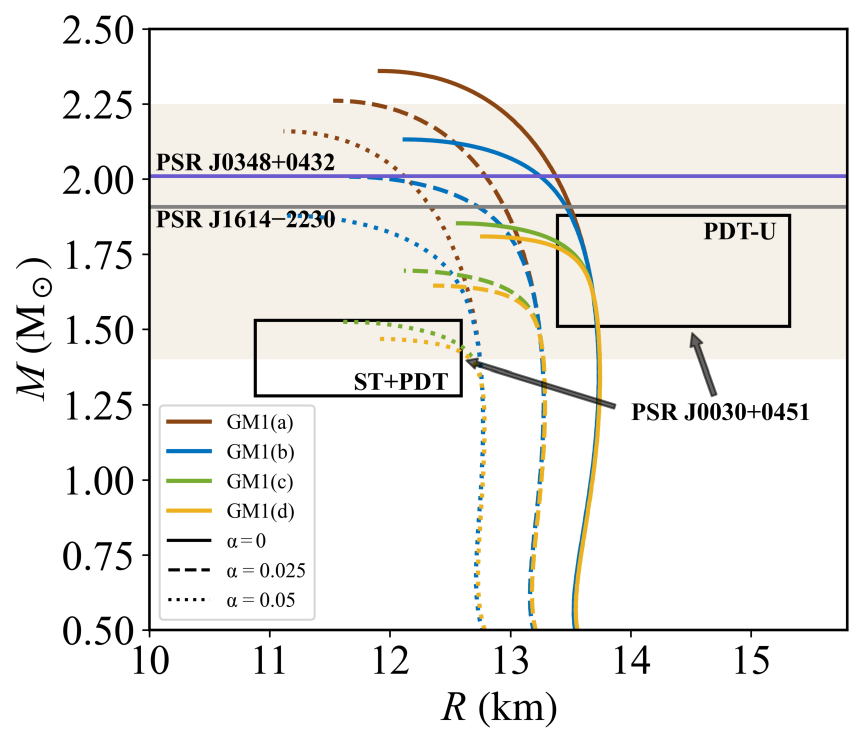}
        }\hfill
    \subfloat{
        \includegraphics[width=3.2 in,height=2.8in]{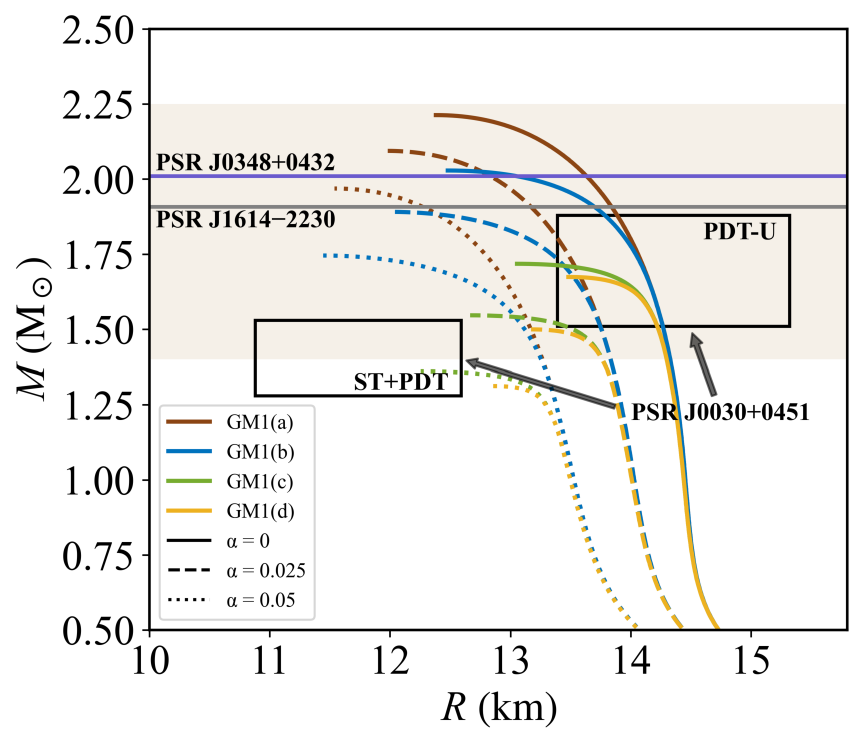}
        }
    \caption{Mass-radius relationships of the traditional neutron stars and the hyperon stars with $\alpha$ = 0, 0.025 and 0.05 under the GM1 and TM1 parameter sets. The shaded part and the colored lines represent the mass range of 1.4-2.25 $M_\odot$\textsuperscript{\cite{fan2024maximum}} as well as the mass and radius measurements of PSR J1614-2230\textsuperscript{\cite{arzoumanian2018nanograv}}, PSR J0348+0432\textsuperscript{\cite{antoniadis2013massive}} and PSR J0030+0451\textsuperscript{\cite{vinciguerra2024updated}}, respectively.}
\end{figure}

\begin{figure}%
    \centering
    \captionsetup{labelfont=bf,labelsep=period}
    \subfloat{
        \includegraphics[width=3.2in,height=2.8in]{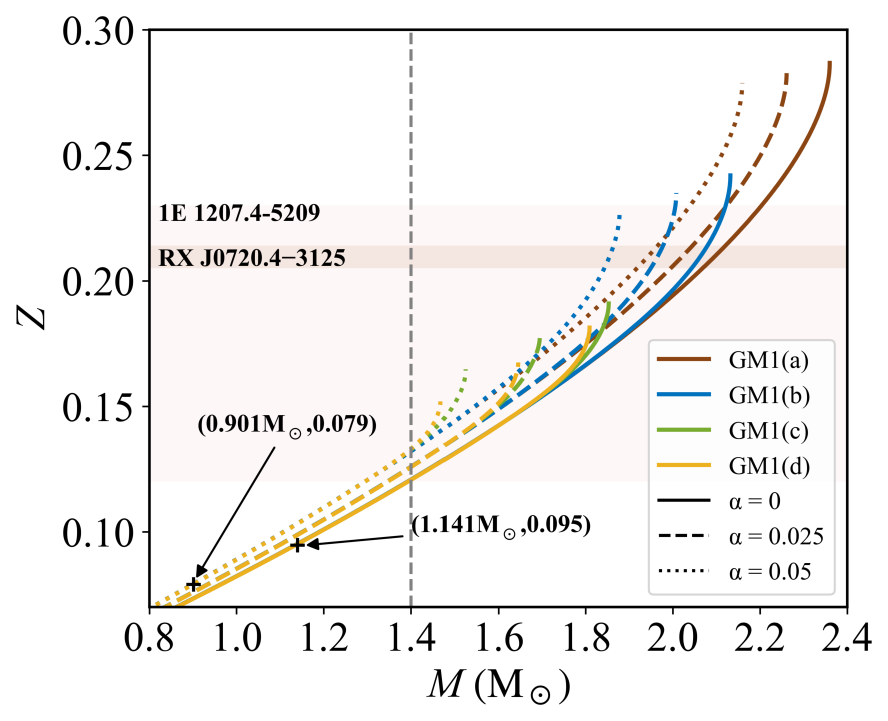}
        }\hfill
    \subfloat{
        \includegraphics[width=3.2in,height=2.8in]{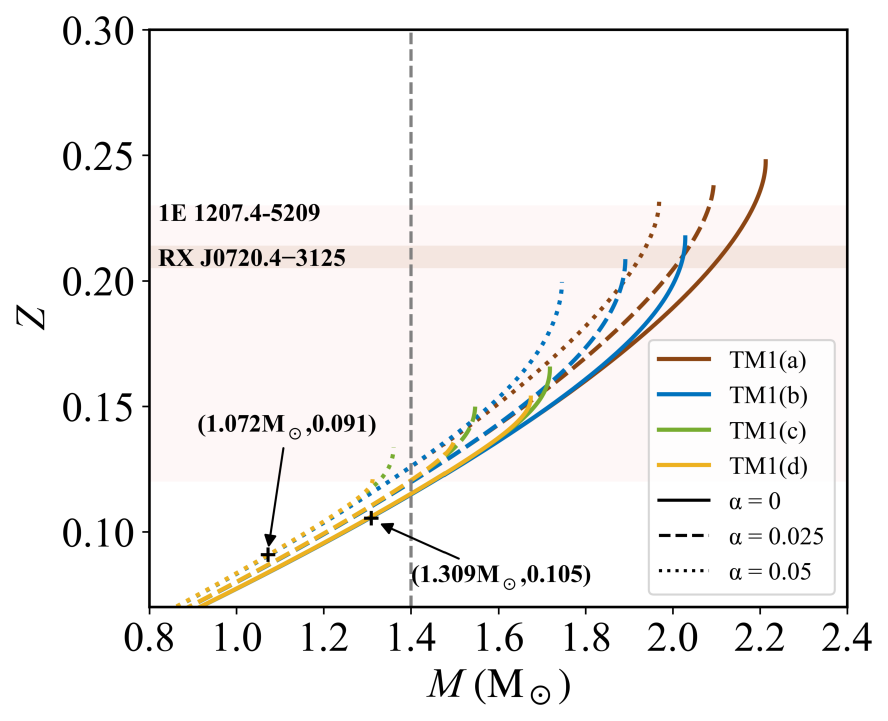}
        }
    \caption{The relationships between gravitational redshift and mass for the traditional neutron stars and the hyperon stars with $\alpha$ = 0, 0.025 and 0.05 under GM1 and TM1 parameter sets. The shaded parts are constrainted by the surface gravitational redshift measurements of 1E 1207.4-5209\textsuperscript{\cite{sanwal2002discovery}} and RX J0720.4−3125\textsuperscript{\cite{hambaryan2017compactness}}, respectively. The vertical dashed line represents a neutron star at the mass of 1.4 $M_\odot$. }

    \captionsetup{labelfont=bf,labelsep=period}
    \subfloat{
        \includegraphics[width=3.2in,height=2.8in]{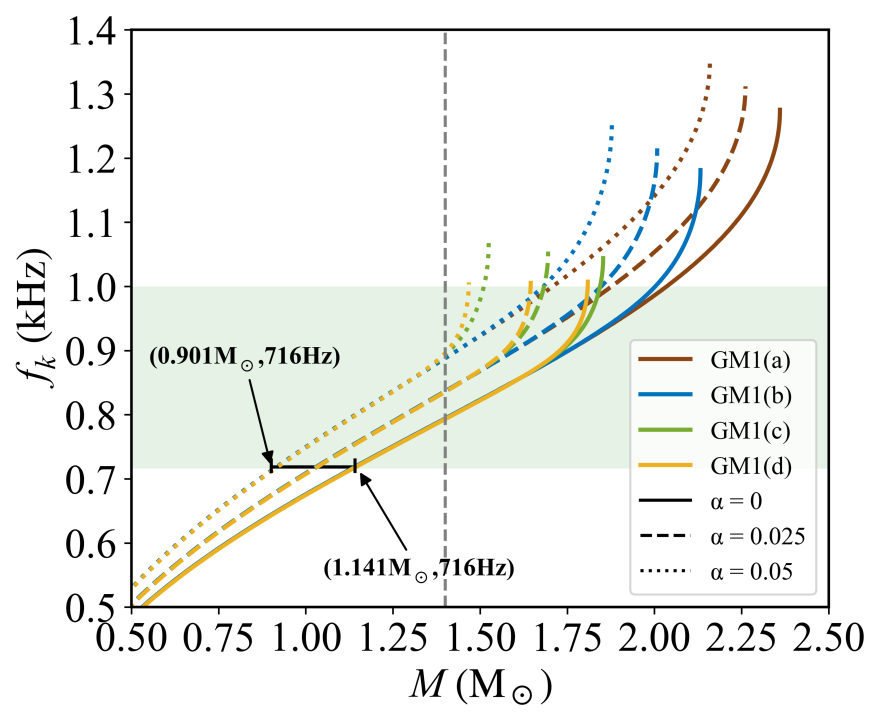}
        }\hspace{0.5mm}
    \subfloat{
        \includegraphics[width=3.2in,height=2.8in]{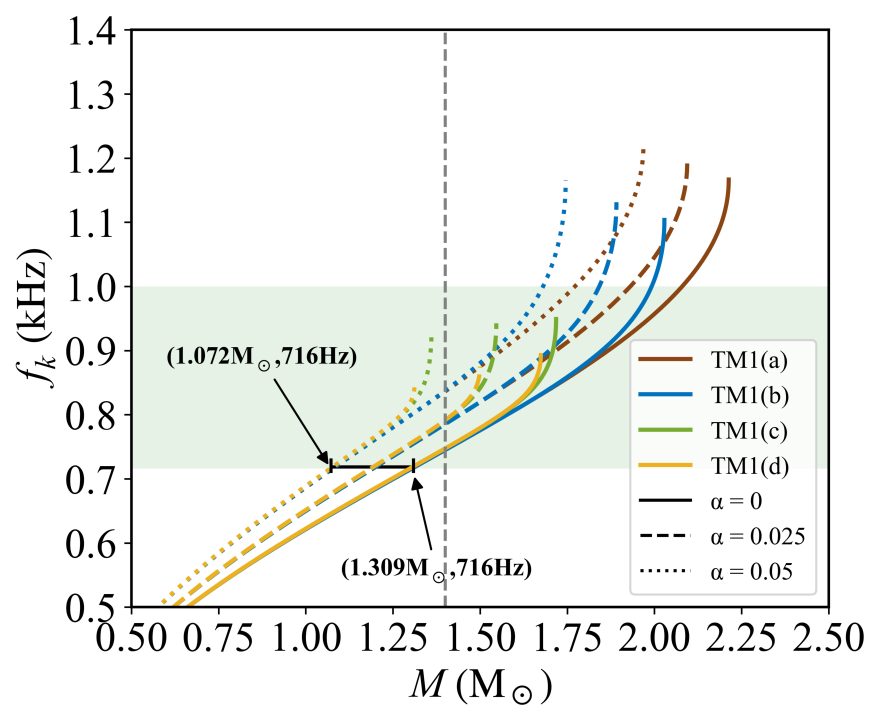}
        }
    \caption{The relationships between Keplerian frequency and mass for the traditional neutron stars and the hyperon stars with $\alpha$ = 0, 0.025 and 0.05 under the GM1 and TM1 parameter sets. The green shaded part indicates that the frequency is in the range of 716-1000 Hz.}
\end{figure}

The EOSs for the traditional neutron stars and the hyperon stars mixing dark energy are shown in Figure 1. With the attendance of dark energy, the EOSs are softened and the degree of softening becomes more pronounced as the fraction of dark energy increases in both the traditional neutron stars and the hyperon stars. After solving the Eqs.(8) and (9) for the traditional neutron stars and the hyperon stars with dark energy combining the mass and radius measurements from the astronomical observations of PSRs J1614-2230 (1.908 $\pm$ 0.016 $M_\odot$)\textsuperscript{\cite{arzoumanian2018nanograv}}, J0348+0432 (2.01 $\pm$ 0.04 $M_\odot$)\textsuperscript{\cite{antoniadis2013massive}} and J0030+0451 ($1.40^{+0.13}_{-0.12}$ $M_\odot$, $11.71^{+0.88}_{-0.83}$ km for ST+PDT model and $1.70^{+0.18}_{-0.19}$ $M_\odot$, $14.44^{+0.88}_{-1.05}$ km for PDT-U model)\textsuperscript{\cite{vinciguerra2024updated}}, the mass–radius relationships are shown in Figure 2. It can be seen that the radius of a given mass traditional neutron star (or hyperon star) decreases with the increase of the fraction of dark energy, which enhances the compactness of the traditional neutron star (or hyperon star). Furthermore, in a dark energy environment, the mass-radius relationships of the traditional neutron stars and the hyperon stars are consistent with the observed mass and radius ranges of PSRs J1614-2230, J0348+0432 and J0030+0451. Especially, the numerical results for the stars with a mass range of 1.35-1.53 $M_\odot$ can satisfy the mass ($1.40^{+0.13}_{-0.12}$ $M_\odot$) and radius ($11.71^{+0.88}_{-0.83}$ km) constraints of PSR J0030+0451 only in the case of involving of dark energy. The theoretical values of the maximum mass and the corresponding radius, the surface gravitational redshift and the Keplerian frequency of the traditional neutron and the hyperon stars are listed in Table 2. As illustrated in Figure 2 and Table 2, the emergence of dark energy causes the maximum mass of the traditional neutron star in the GM1 (TM1) parameter set to decrease from 2.360 $M_\odot$ (2.213 $M_\odot$) to 2.159 $M_\odot$ (1.969 $M_\odot$), and the maximum mass of the hyperon star decrease from 2.132 $M_\odot$ (2.029 $M_\odot$) to 1.525 $M_\odot$ (1.361 $M_\odot$).

Figure 3 shows the relationships between the surface gravitational redshift and the mass for the traditional neutron stars and the hyperon stars with the different fraction of dark energy. The two colored shaded regions represent the ranges of the surface gravitational redshift measurements for 1E 1207.4-5209 ($z=0.12-0.23$)\textsuperscript{\cite{sanwal2002discovery}} and RX J0720.4-3125 ($z =0.205^{+0.006}_{-0.003}$)\textsuperscript{\cite{hambaryan2017compactness}}, respectively. The theoretical values of the surface gravitational redshift and the Keplerian frequency for 1.4 $M_\odot$ are also listed in Table 2. As seen in Figure 3, the surface gravitational redshift increases as the growing fraction of dark energy in a traditional neutron star (or hyperon star) at the same mass. Moreover, the numerical results in Figure 3 under the SU(3) flavor symmetry can simultaneously satisfy the observed constraints of 1E 1207.4-5209 and RX J0720.4-3125.

Figure 4 displays the relationships between the Keplerian frequency and the mass for the traditional neutron stars and the hyperon stars with the different fraction of dark energy combining with the observed frequency limits of PSR J1748-2446ad. PSR J1748-2446ad is the currently known fastest rotating pulsar with a spin frequency of $f=716$ Hz, which was discovered in a globular cluster Terzan 5 in 2006\textsuperscript{\cite{hessels2006radio}}. In fact, many researchers have done a great deal of studies on the formation mechanism and the possible detection methods of sub-millisecond pulsars\textsuperscript{\cite{han2004searching,kaaret2007evidence,du2009formation,bult2021x}}. Therefore, we take $f=1000$ Hz as the upper limit value of frequency\textsuperscript{\cite{xu2023probing}} to illustrate the numerical results in Figure 4. From Figure 4, we can see that the Keplerian frequency of the traditional neutron star (or hyperon star) increases with the increase of the fraction of dark energy under the same mass conditions. Especially, as shown in Figure 4 and Table 2, when the fraction of dark energy is relatively high, the low and medium-mass traditional neutron star (or hyperon star) can also reach $f=1000$ Hz. Furthermore, we infer that PSR J1748-2446ad is very likely a low-mass pulsar, regardless of whether PSR J1748-2446ad contains dark energy. Specifically, the mass and surface gravitational redshift of PSR J1748-2446ad without dark energy under the GM1 (TM1) parameter set are 1.141 $M_\odot$ (1.309 $M_\odot$) and 0.095 (0.105), respectively. The corresponding values for the GM1 (TM1) parameter set are 0.901 $M_\odot$ (1.072$M_\odot$) and 0.079 (0.091) for PSR J1748-2446ad with dark energy and $\alpha=0.05$. If the observational data about PSR J1748-2446ad can provide further information in the future, the above results will be beneficial for exploring its composition.

\begin{table}[htbp]
\tabcolsep=15pt
\small
\captionsetup{labelfont=bf,labelsep=space}
\caption{Maximum mass $M_{max}$ and the corresponding radius $R$, the surface gravitational redshift $Z$ and the Keplerian frequency $f_K$ of the traditional neutron and the hyperon stars with $\alpha=$ 0, 0.025 and 0.05 under GM1 and TM1 parameter sets. $Z_{1.4}$ and $f_{K,1.4}$ represent the theoretical values of the surface gravitational redshift and Keplerian frequency at the mass of $1.4M_\odot$.}
\vglue5pt
\resizebox{\linewidth}{!}{
\begin{tabular}{cccccccc}
        \toprule
         &$\alpha$ & $R$(km)  & $M_{max}(M_\odot)$ & $Z$  & $f_K$(kHz) & $Z_{1.4}$ & $f_{K,1.4}$(kHz) \\
        \midrule
        \text{GM1(a)}& 0 & 11.917 & 2.360 & 0.287 & 1.181 & 0.122 & 0.796  \\
         \textbf{ } &0.025 & 11.529 & 2.261 & 0.283 &1.215 & 0.126 &0.838\\
         \textbf{ } &0.05 & 11.116& 2.159 & 0.279 & 1.254 & 0.133 & 0.890\\
         
         \midrule
         \text{TM1(a)} &0 & 12.379 & 2.213 & 0.248 &1.167 & 0.115 & 0.745\\
         \textbf{ }&0.025 & 11.978 & 2.094 &0.240 &1.192 & 0.120 & 0.787 \\
         \textbf{ }& 0.05 & 11.544 & 1.969 & 0.232 & 1.222& 0.126 & 0.838\\

         \midrule
         \text{GM1(b)} &0& 12.124 & 2.132 & 0.242 & 1.094 & 0.122 & 0.796 \\
         \textbf{ }&0.025& 11.662 & 2.008&0.235 & 1.125 & 0.126 & 0.838  \\
         \textbf{ }&0.05&11.147 & 1.878 & 0.228 &1.164 &0.132  &0.889 \\
         \midrule
         \text{TM1(b)}  &0&12.474 & 2.029& 0.217 &1.104 & 0.115 & 0.745  \\
         \textbf{ }&0.025&  11.990 & 1.891 & 0.209&1.131&0.120 & 0.787 \\
         \textbf{ }&0.05& 11.446 & 1.746 & 0.200 & 1.165 & 0.126 & 0.838\\

          \midrule
         \text{GM1(c)} & 0 &12.563 & 1.853& 0.191 & 0.967 & 0.121 & 0.794  \\
         \textbf{ }&0.025 &12.117 & 1.695 & 0.178&0.976& 0.126 & 0.835   \\
         \textbf{}&0.05 &11.598 & 1.525 & 0.165 &0.989  & 0.133 & 0.894  \\
         \midrule
         \text{TM1(c)}  & 0 &  13.054 & 1.718 & 0.165 &0.949 & 0.115 & 0.748  \\
         \textbf{ }&0.025&12.664 &1.546 & 0.150 &0.942 & 0.121 & 0.792   \\
         \textbf{}&0.05 & 12.257 & 1.361 & 0.134 & 0.928 &- &- \\

         \midrule
         \text{GM1(d) }  &0 &  12.764& 1.809& 0.181& 0.933 & 0.121 & 0.794 \\
         \textbf{ }&0.025 &12.356 & 1.645 & 0.167 &0.934 & 0.126 & 0.836 \\
         \textbf{ }&0.05 &11.913 & 1.468 & 0.152 & 0.932 & 0.133 & 0.897 \\

         \midrule
         \text{TM1(d) }  &0 &  13.481 & 1.674 & 0.153 & 0.893 &0.115 & 0.748  \\
         \textbf{}&0.025 & 13.171 & 1.500 & 0.138 & 0.848 &0.121 & 0.792 \\
         \textbf{ }&0.05 & 12.856 & 1.312 & 0.121 & 0.875 & - &- \\

 \bottomrule \end{tabular} \small
}

\end{table}

\section{Conclusions}

In the work, we investigate the influence of dark energy on the main macroscopic properties of the traditional neutron and the hyperon stars using the GM1 and TM1 parameter sets under the SU(6) spin-flavor and the SU(3) flavor symmetries within the framework of relativistic mean field theory. The results illustrate that the inclusion of dark energy in the traditional neutron stars and the hyperon stars lead to softening of EOSs and enhancement of compactnesses. For the traditional neutron or hyperon stars with the same mass, the theoretical values of the surface gravitational redshift and the Keplerian frequency go up with the dark energy fraction increases. The increase of the Keplerian frequency provides a possible way to understand and explain the glitch phenomenon of some pulsars. The numerical results of the traditional neutron stars or the hyperon stars containing dark energy are in good agreement with the corresponding observations of PSR J1614-2230, PSR J0348+0432, PSR J0030+0451, RX J0720.4-3125 and 1E 1207.4-5209. In addition, we also infer that the mass and surface gravitational redshift value ranges of PSR J1748-2446ad with and without dark energy. For PSR J1748-2446ad without dark energy, its mass and surface gravitational redshift under the GM1 (TM1) parameter set are 1.141 $M_\odot$ (1.309 $M_\odot$) and 0.095 (0.105), respectively. For PSR J1748-2446ad with dark energy and $\alpha=0.05$, the corresponding values of the GM1 (TM1) parameter set are 0.901 $M_\odot$ (1.072$M_\odot$) and 0.079 (0.091), respectively. It indicates that if PSR J1748-2446ad contains dark energy, its mass would be very low. In the future, more observed data will be used to constrain the EOSs of neutron stars and dark energy with the continuous progress of astronomical observations, which are very beneficial for revealing the composition of neutron stars.

\section*{Acknowledgements}
The authors sincerely thank Pro. J. L. Han from the National Astronomical Observatories, Chinese Academy of Sciences for his valuable and constructive comments on the researches of neutron star theory and pulsar observation, which enable us to have more profound and comprehensive understanding of the field. This work is partially funded by the Horizontal Scientific Research Project of the National Astronomical Observatories of CAS (Grant No. E0900501) and the Theoretical Fundamental Research Special Project of the Changchun Observatory, National Astronomical Observatories, CAS (Grant No.Y990000205).

\bibliographystyle{unsrt}
\bibliography{CTP-revised}
\end{CJK}
\end{document}